\documentclass{article}
\usepackage{spconf}
\usepackage[T1]{fontenc}
\usepackage[latin1]{inputenc}
\usepackage{graphicx}
\usepackage{subfigure}
\usepackage{color}
\usepackage{mathrsfs}
\usepackage{amsfonts}
\usepackage{amsmath}
\newcommand{\avg}[1]{\left< #1 \right>} 

\newtheorem{mydef}{Definition}[section]

\begin{document}

\def\x{{\mathbf x}}
\def\L{{\cal L}}

%
\title{Model-free control of microgrids }
\name{Loïc Michel$^{\ast}$, Wim Michiels$^{\ast}$ and Xavier Boucher$^{\dag}$}

\address{$^{\ast}$ Department of Computer Science \\
KU Leuven \\
Celestijnenlaan 200A \\
B - 3001 Heverlee \\
E-mail: $\{$Loic.Michel, Wim.Michiels$\}$@cs.kuleuven.be \\
$^{\dag}$ E-mail: xavier.b.eng@gmail.com}
%
%
%

%
\maketitle
\begin{abstract}
A new "model-free" control methodology is applied for the first time to power systems included in microgrids networks. We evaluate its performances
regarding output load and supply variations in different working configuration of the microgrid. Our approach, which utilizes "intelligent" PI controllers, does not require any converter or microgrid model identification while ensuring the stability and the robustness of the controlled system. Simulations results show that with a simple control structure, the proposed control method is almost insensitive to fluctuations and large load variations.
\end{abstract}
\section{Introduction}
\label{sec:intro}
The model-free control methodology, originally proposed by \cite{esta}, has been widely successfully applied to many mechanical and electrical processes. The model-free control provides good performances in disturbances rejection and an efficient robustness to the process internal changes. A preliminary work on power electronics \cite{Michel} presents the successful application of the model-free control method to the control of dc/dc converters. The control of inverter-based microgrids has been deeply studied and some advanced methods have been successfully developed and tested (e.g. \cite{Yang} \cite{Mohamed} \cite{Bal}).  This paper extends the previous results to the control of inverter-based microgrids in different situations related to islanded and grid-connected modes. In particular, we will show that the proposed control method is robust to strong load variations either in voltage, current or power control cases.

The paper is structured as follows. Section II presents an overview of the model-free control methodology including its advantages in comparison with classical methodologies. Section III discusses the application of the model-free control to inverters. Some concluding remarks may be found in Section IV.

\section{Model-free control: a brief overview}\label{mfc}
\subsection{General principles}

We only assume that the plant behavior is well approximated in its
operational range by a system of ordinary differential equations,
which might be highly nonlinear and time-varying. The system, which is SISO,
may be therefore described by the input-output equation:

\begin{equation}\label{es}
E (t, y, \dot{y}, \dots, y^{(\iota)}, u, \dot{u}, \dots, u^{(\kappa)}) = 0
\end{equation}

\begin{itemize}
\item $u$ and $y$ are the input and output variables,
\item $E$, which might be unknown, is assumed to be a
sufficiently smooth function of its arguments.
\end{itemize}

From (\ref{es}), we define an {\it ultra-local} model, which represents (\ref{es}) over a small time period.

\begin{mydef}\label{mydef-modele_F}
\cite{esta} If $u$ and $y$ are respectively the variables of input and output of a system to be controlled, then this system can be described as the ultra-local model defined by:
\begin{equation}\label{mod}
y^{(n)} = F + \alpha u
\end{equation}
where
$\alpha \in \mathbb{R}$ is a {\em non-physical} constant parameter,
such that $F$ and $\alpha u$ are of the same magnitude,
and $F$ contains all structural information of the process.
\end{mydef}
In all the numerous known examples, it was possible to set $n = 1$ or $2$ \cite{Fliess_Mar}. Let us emphasize that one only needs to give an approximate numerical value to $\alpha$. The gained experience shows that taking $n = 2$ allows to stabilize switching systems.

\subsection{Intelligent PI controllers}

\begin{mydef}
\cite{esta} \label{mydef-iPI} We close the loop via the {\em intelligent PI controller}, or {\em i-PI} controller,
\begin{equation}\label{eq:ipi}
u = - \frac{[F]}{\alpha} + \frac{{y}^{(n) \, \ast}}{\alpha}  + \mathcal{C}(\varepsilon)
\end{equation}
where
\begin{itemize}
\item $[F]$ is an estimate of $F$ in (\ref{mod}), computed on-line as $[y^{(n)}]-\alpha u$, where $[y^{(n)}]$ is an approximation of the output derivative;
\item $y$ is the measured output to control and $y^\ast$ is the output reference trajectory;
\item $\varepsilon = y^\ast - y$ is the tracking error;
\item $\mathcal{C}(\varepsilon)$ is of the form $K_p \varepsilon + K_i \int \varepsilon$. $K_p$, $K_i$ are the usual tuning gains.
\end{itemize}

Equation (\ref{eq:ipi}) is called the model-free control law or model-free law.
\end{mydef}

The i-PI controller (\ref{eq:ipi}) is compensating the poorly known term $F$ and controlling the system therefore boils down to the control of an integrator. The tuning of the gains $K_P$ and $K_I$ becomes therefore straightforward.

Our implementation of (\ref{eq:ipi}) assumes a sampled-data control context, where the control input is kept constant over the inter-sampling interval and the output derivatives are approximated by finite-differences of the outputs. At the $k$th sampling instants, we have \cite{Michel}:
\begin{multline}\label{eq:ipi2}
u_k = u_{k-1} - \frac{1}{\alpha T_c^2} \left\{ \left( y_{k-1} - 2 y_{k-2} + y_{k-3} \right) \right. -   \\
\left. \left( y^*_{k-1} - 2 y^*_{k-2} + y^*_{k-3} \right) \right\} + C(y_{k-1}^*-y_{k-1})
\end{multline}

\noindent
where $u_k$ refers to the averaged duty-cycle at the $k$th  sampling instant and $T_c = 0.1$ ms is the switching period. The main advantage of the proposed control approach is that sudden changes in the model, e.g. due to load changes, and model uncertainty can be overcome as $F$ in (\ref{mod}) is re-estimated at every sampling instant from the output derivatives and inputs. We note that the potential amplification of noise by differentiation of the output can be countered by using moving average filters, see \cite{fliess}.


To illustrate the utilization of the model-free control in a microgrid environment, the following results present the simulation of a voltage-controlled inverter, a tri-phase controlled inverter and a power controlled inverter under disturbances such as e.g. load changes. We compare the results with a PI control that has been tuned using an ITAE criteria in order to optimize the transient with the initial load \cite{Awouda}. Simulations have been performed using the averaging method \cite{Sun} \cite{sigma} for which the controlled inputs in every case correspond to the averaged duty-cycle values that drive each IGBT.

\section{Control of inverter-based microgrid}

\subsection{Voltage-controlled inverter}

We apply in this section the proposed method to the control of the output voltage of inverters, which are used in typical configurations within microgrid \cite{Lopes} in both stand-alone mode and grid-connected mode. All the inductors and capacitors described on the schemes have their values respectively close to 1 mH and 10 $\mu$F. The dc bus voltage $E$ is equal to 400 V and we take $\alpha = 30$ in (\ref{mod}).

\subsubsection{Single load}

\begin{figure}[!h]
\centering
\includegraphics[width=8cm]{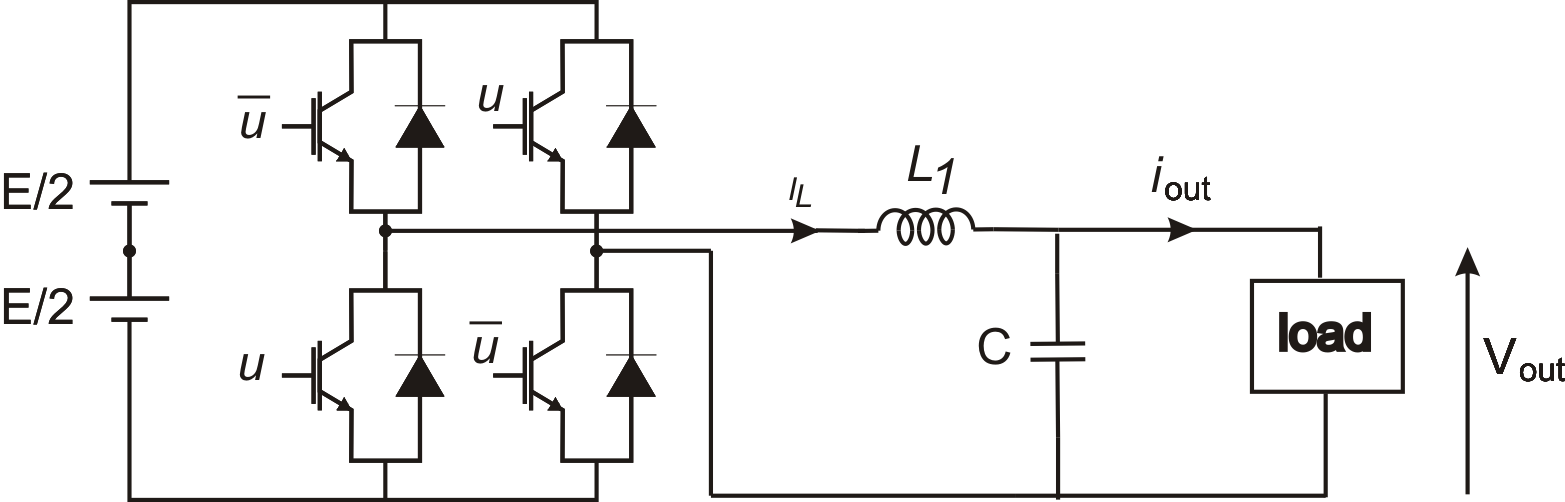}
\caption{Full bridge inverter with a load.}
\label{fig:fig_0}
\end{figure}

\begin{figure}[!h]
\centering
\subfigure[\footnotesize i-PI control]{\includegraphics[width=8cm]{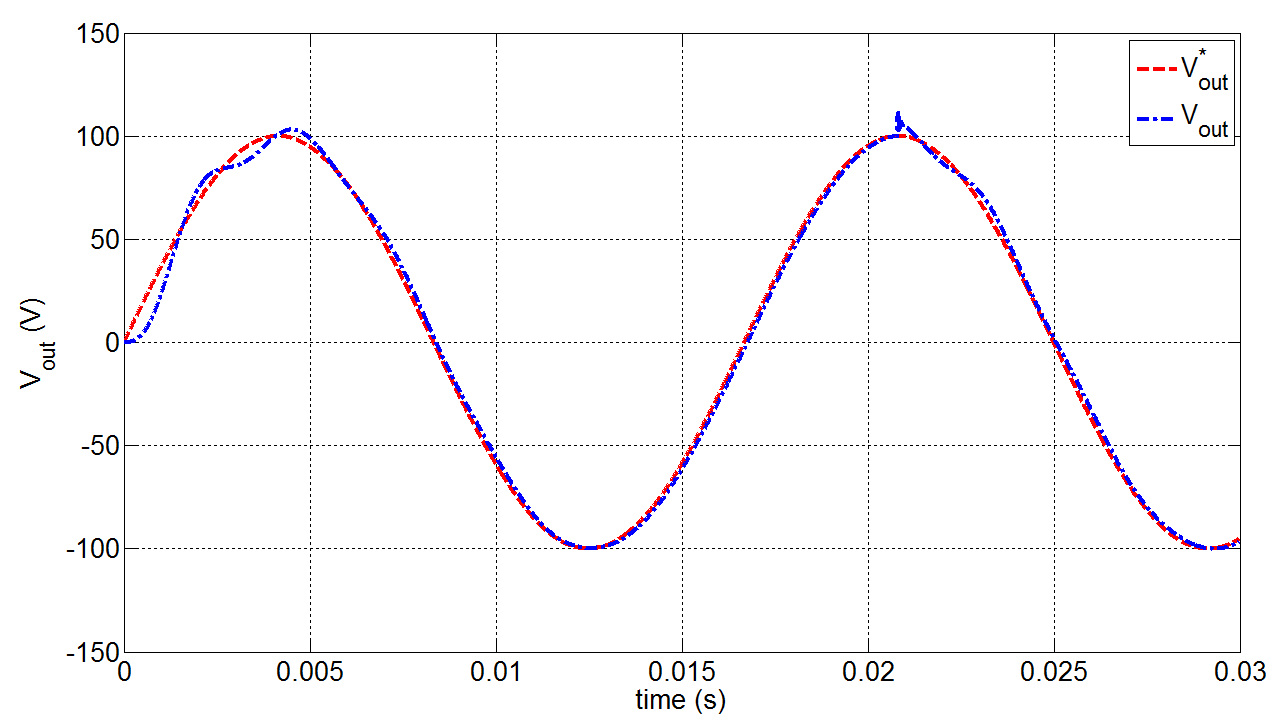}\label{fig:fig_2_iPI}}
\subfigure[\footnotesize PI control]{\includegraphics[width=8cm]{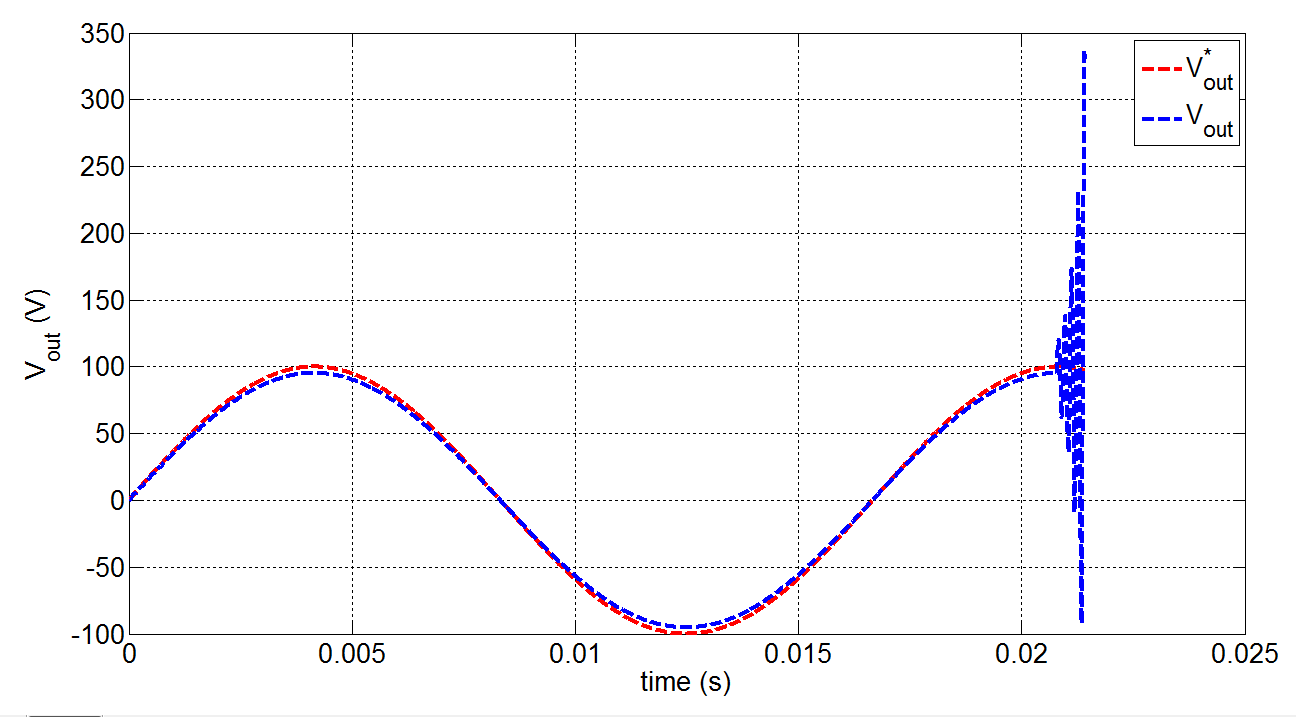}\label{fig:fig_2_PI}}
\caption{Comparison between the PI and i-PI control ($K_p = 20, \, K_i = 0$) for the voltage-controlled inverter.}
\label{fig:fig_2}
\end{figure}

Consider a single-phase inverter working in stand-alone mode, driven by the duty-cycle $u$, for which the output voltage $v_{out}$ is controlled (Fig. \ref{fig:fig_0}). The load is a resistor $R$ that switches from $R \approx 10 \, \Omega$ to $R \approx 1000 \, \Omega$ at $t = 0.02$ s. Figure \ref{fig:fig_2} 
presents the output voltage response of the inverter according to the output voltage reference $v_{out}^*$ when a classical PI controller and an i-PI controller are considered.

%

\subsubsection{Multiple loads}

\begin{figure}[!h]
\centering
\includegraphics[width=8cm]{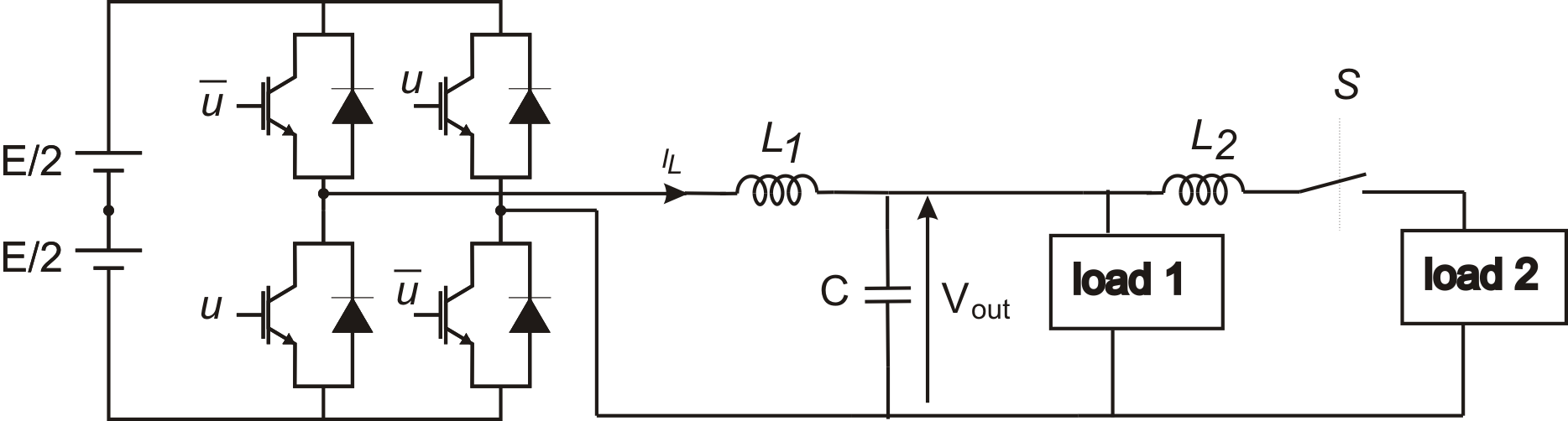}
\caption{Full bridge with two loads.}
\label{fig:fig_4}
\end{figure}

Consider a single-phase inverter working in stand-alone mode, driven by the duty-cycle $u$, for which the output voltage $v_{out}$ is controlled (Fig. \ref{fig:fig_4}).

\begin{figure}[!h]
\centering
\includegraphics[width=8cm]{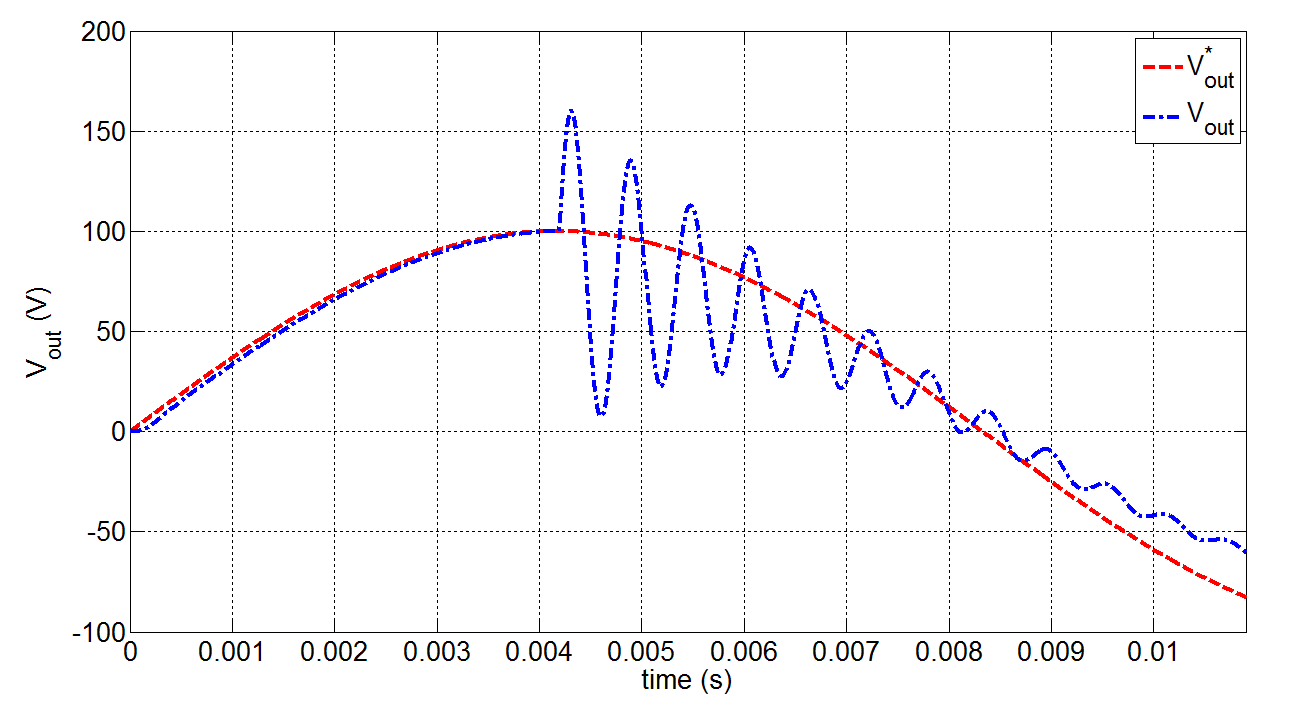}
\caption{Response of the inverter in open-loop (the second load "2" is connected at $t = 0.0042$ s).}
\label{fig:fig_5}
\end{figure}

\begin{figure}[!h]
\centering
\subfigure[\footnotesize $K_p = 20$, $K_i = 0$]{\includegraphics[width=8cm]{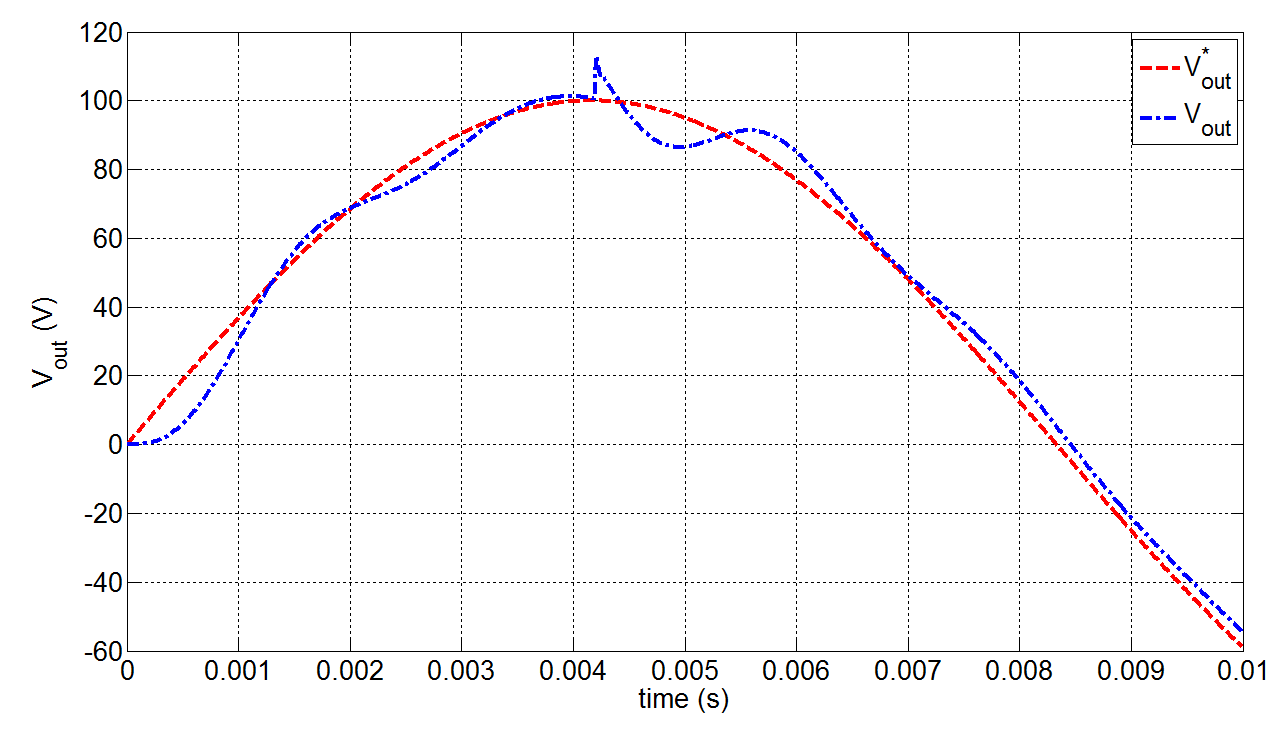}\label{fig:fig_6a}}
\subfigure[\footnotesize $K_p = 50$, $K_i = 100$]{\includegraphics[width=8cm]{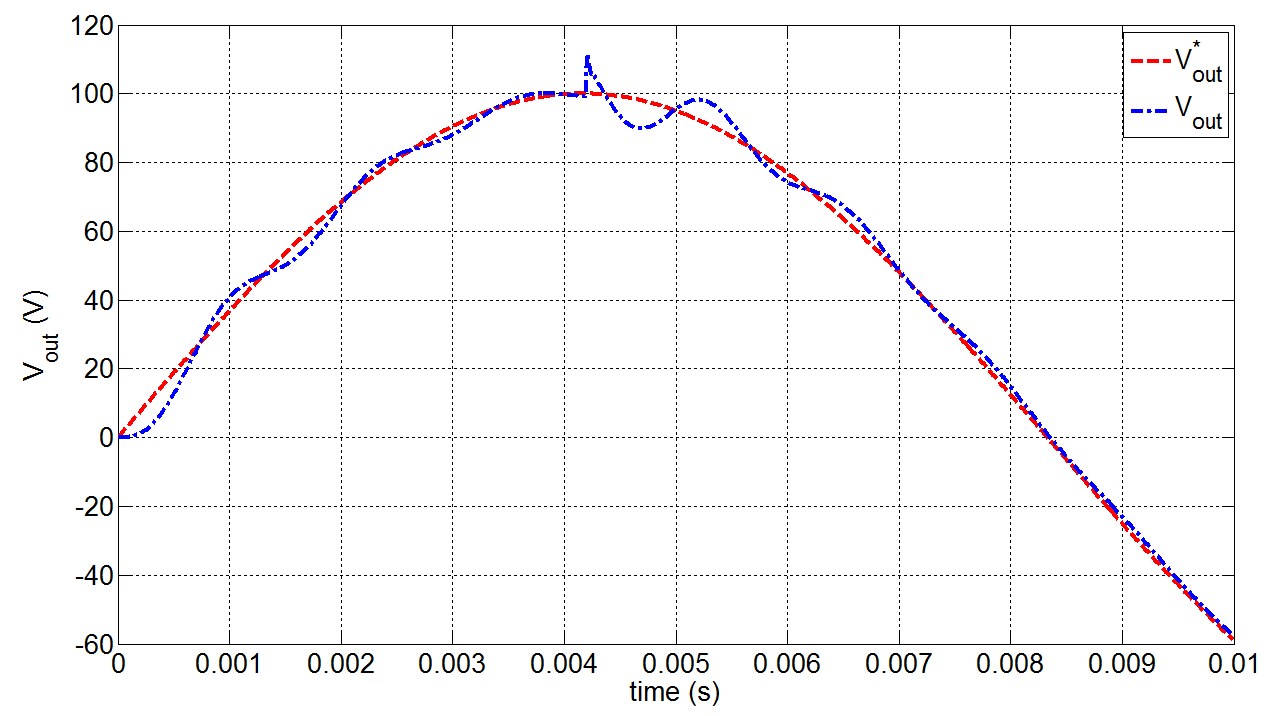}\label{fig:fig_7a}}
\caption{i-PI control of the output voltage.}
\label{fig:fig_7}
\end{figure}

The inverter is firstly loaded by a resistor (load "1") and then a second {\it unknown} load (load "2") is added at $t = 0.0042$ s. Figure \ref{fig:fig_5} shows the output voltage response of the inverter in open-loop. 
Figure \ref{fig:fig_7} presents the inverter output voltage $v_{out}$ response with an i-PI controller for different $K_p$ and $K_i$ parameters.

\subsection{Tri-phase current-controlled inverter}

\begin{figure}[!h]
\centering
\includegraphics[width=8cm]{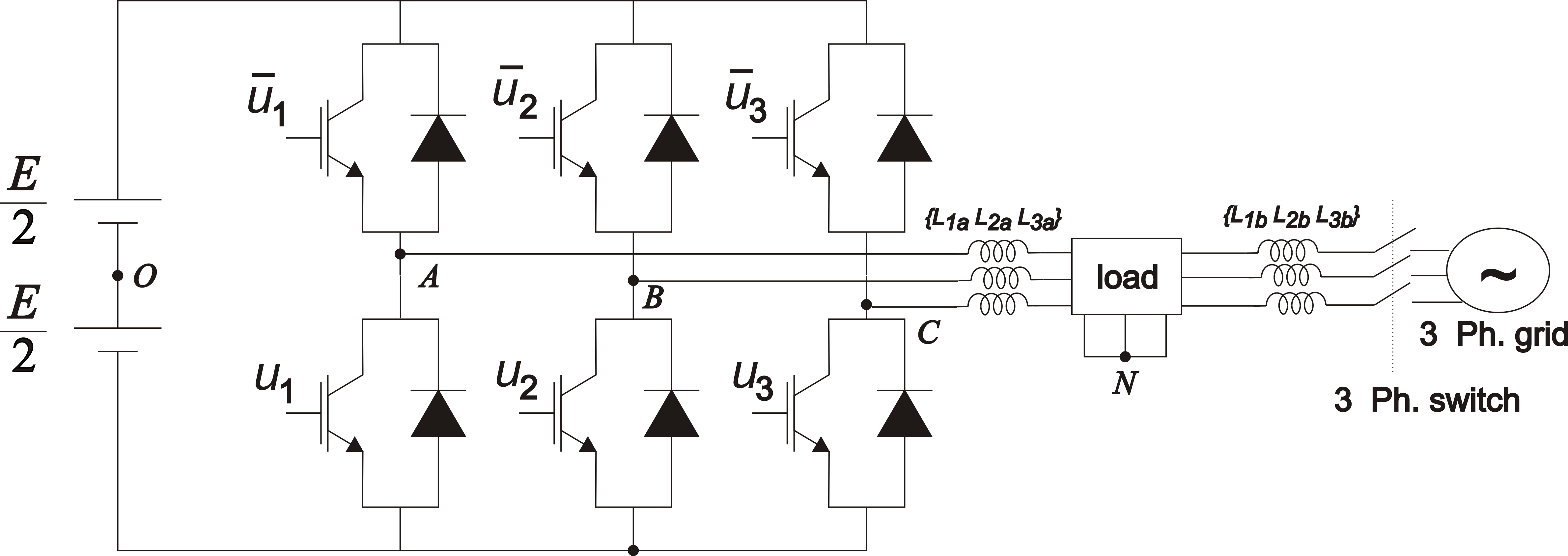}
\caption{Tri-phase bridge inverter connected to the grid.}
\label{fig:fig_10}
\end{figure}

Consider a current-controlled tri-phase inverter working in both stand-alone mode / grid-connected mode. The current in each phase is controlled (Fig. \ref{fig:fig_10}) by i-PI (each phase has its own i-PI controller); $I_{L_1 \, a}, I_{L_2 \, a}, I_{L_3 \, a}$ are the controlled currents going through the inductors $L_{1 \, a}$, $L_{2 \, a}$,  $L_{3 \, a}$ and  $I_{L_1 \, a}^*$, $I_{L_2 \, a}^*$, $I_{L_3 \, a}^*$ are the corresponding reference currents. Mathematically we have a multiple input, multiple output system (MIMO) and the local models, each pairing one input and one output, take the form :
\[
\left\{\begin{array}{lll}
\dot y_1&=&F_1(y_1,y_2,y_3,u_2,u_3)+\alpha_1 u_1,\ \ \\
&\vdots&\\
\dot y_3&=&F_3(y_1,y_2,y_3,u_1,u_2)+\alpha_3 u_3,
\end{array}\right.\]
hence the interdependencies between the inputs and outputs that are not-paired are absorbed in the terms $F_i$. The terms $u_1$, $u_2$ and $u_3$ are the averaged duty-cycle that drive the IGBTs of the bridge.

\begin{figure}[!h]
\centering
\includegraphics[width=8cm]{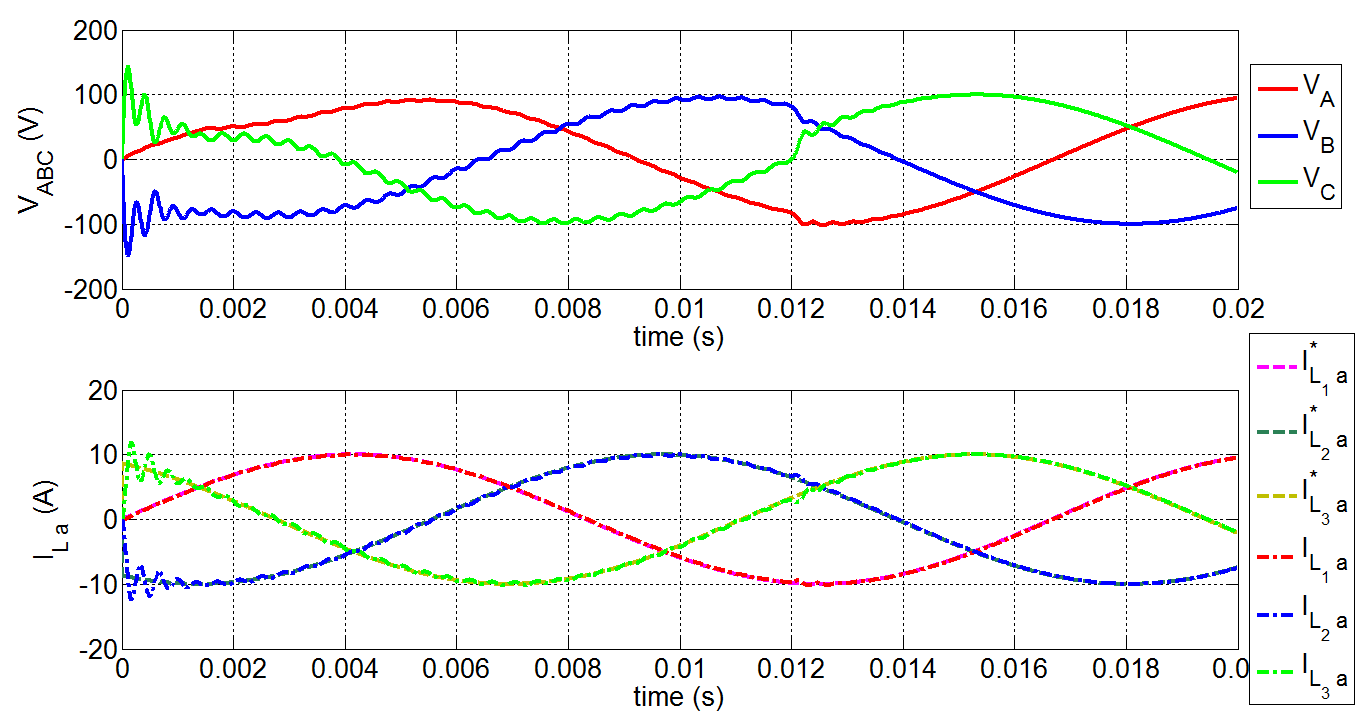}
\caption{Stand-alone mode current-controlled inverter with load changes.}
\label{fig:fig_11}
\end{figure}

The load is composed of a tri-phase resistor ($\approx 10 \, \Omega$) and a tri-phase capacitor ($\approx 10 \, \mu$F). Figure \ref{fig:fig_11} presents the voltages and currents of the inverter in stand-alone mode with a tri-phase load change ($R = 1000 \, \Omega, \, C = 0.1 \mu$F) at $t = 0.012$ s. Results are similar in the case of unbalanced conditions. We take $\alpha_1 $= $\alpha_2 $= $\alpha_3$ = 30, $K_p = 500$ and $K_i = 300$.

\begin{figure}[!h]
\centering
\subfigure[\footnotesize Inverter alone (without control).]{\includegraphics[width=8cm]{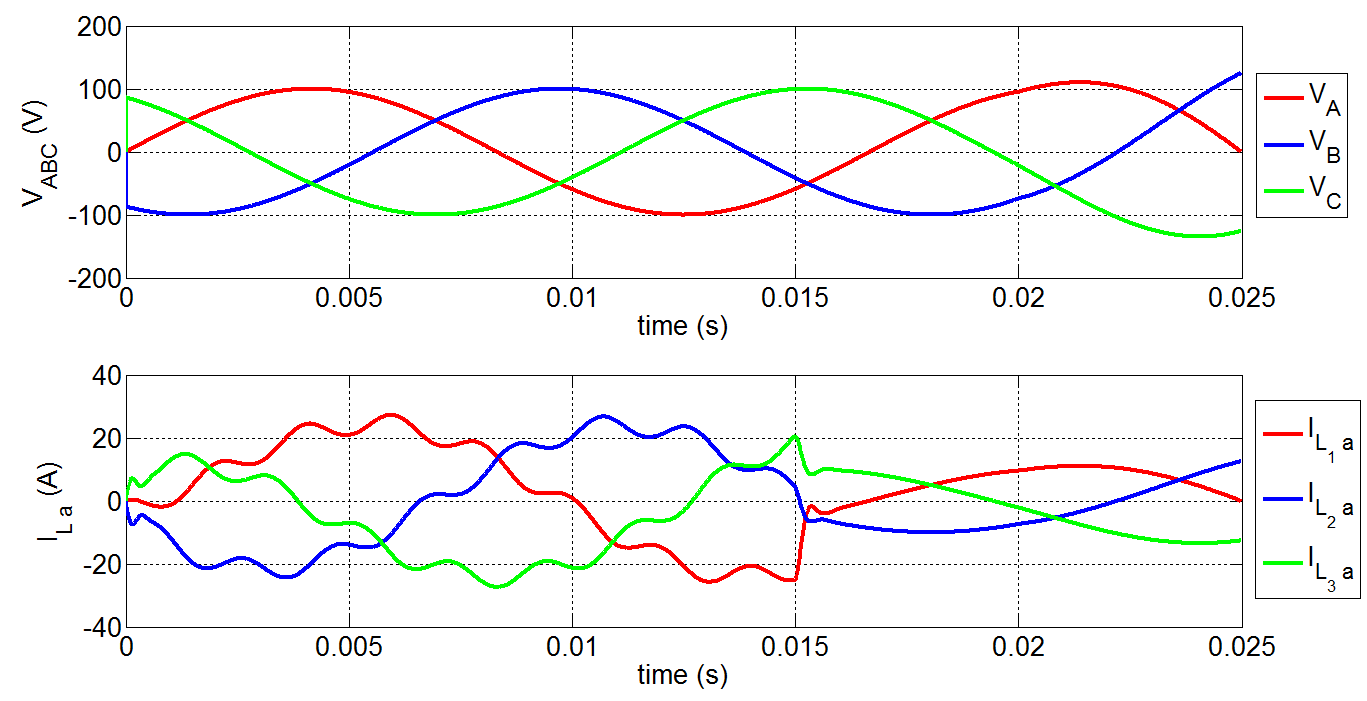}\label{fig:fig_12a}}
\subfigure[\footnotesize i-PI control.]{\includegraphics[width=8cm]{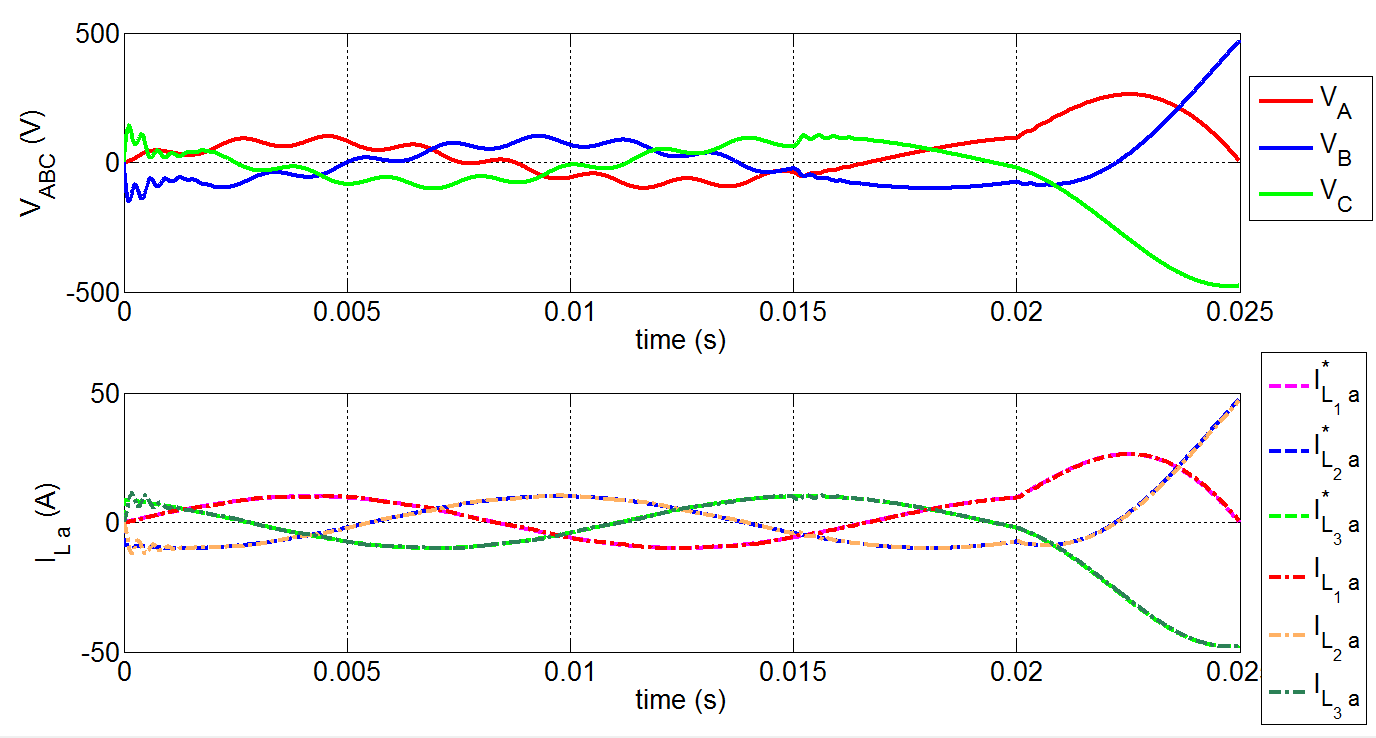}\label{fig:fig_12b}}
\caption{Perturbated grid and disconnection from the grid.}
\label{fig:fig_12}
\end{figure}

A grid disconnection is presented Fig. \ref{fig:fig_12} : a sinusoidal perturbation of 25 $\%$ of the grid amplitude at 500 Hz is added to the grid and the inverter is disconnected from the grid at $t = 0.015$ s.

\subsection{Power-controlled inverter}

Controlling the output power of an inverter is important when considering parallelization of inverters and load sharing \cite{Divan} \cite{Chih}.

\begin{figure}[!h]
\centering
\includegraphics[width=8.5cm]{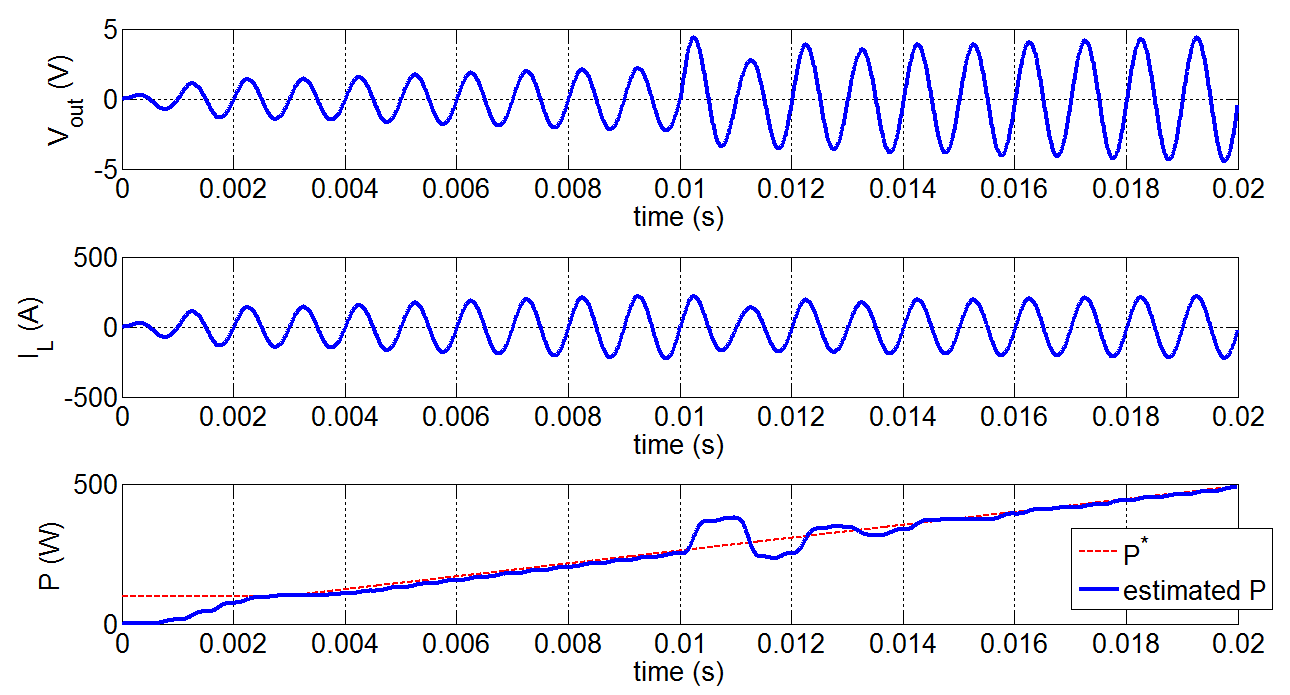}
\caption{Power-controlled inverter.}
\label{fig:fig_14}
\end{figure}

Consider the single-phase full bridge inverter described Fig. \ref{fig:fig_0} working in stand-alone mode; the load is a resistor ($R = 100 \, \Omega$). We consider in this section the control of the active power $\avg{\mathcal{P}}$ at the output of the inverter for which the i-PI controller is configured with $\alpha = 30$, $K_p = 20$ and $K_i = 0$. The output active power $\mathcal{P}$ is defined by :

\begin{equation}
\avg{\mathcal{P}} = \frac{1}{T_c} \int_{t - T_c}^t v_{out} \, i_{out} \, d \, t
\end{equation}

\noindent
and its estimator is based on a moving-average filter. This is a direct control and the i-PI controller corrects the amplitude of the output sinusoidal signal in order to satisfy the power reference $\avg{\mathcal{P}}^*$.  Figure \ref{fig:fig_14} shows the active output estimated power $\avg{\mathcal{P}}$ of the inverter controlled by i-PI. A load change occurs ($R = 50 \, \Omega$) at $t = 0.01$ s. This strategy can also work in tri-phase systems.

\subsection{Parallel inverters}

\begin{figure}[!h]
\centering
\includegraphics[width=7.5cm]{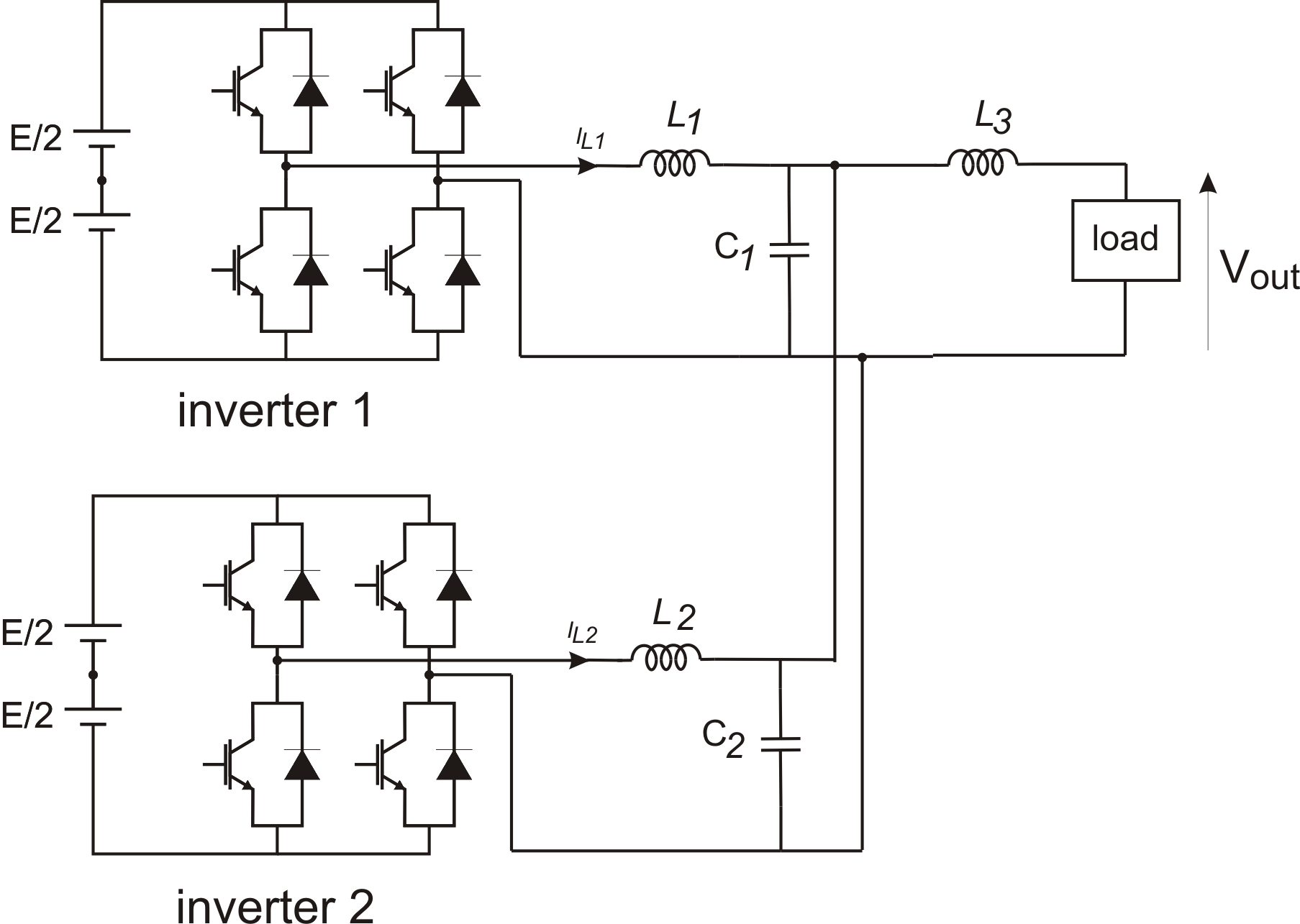}
\caption{Parallelization of inverters.}
\label{fig:fig_16}
\end{figure}

\begin{figure}[!h]
\centering
\includegraphics[width=8cm]{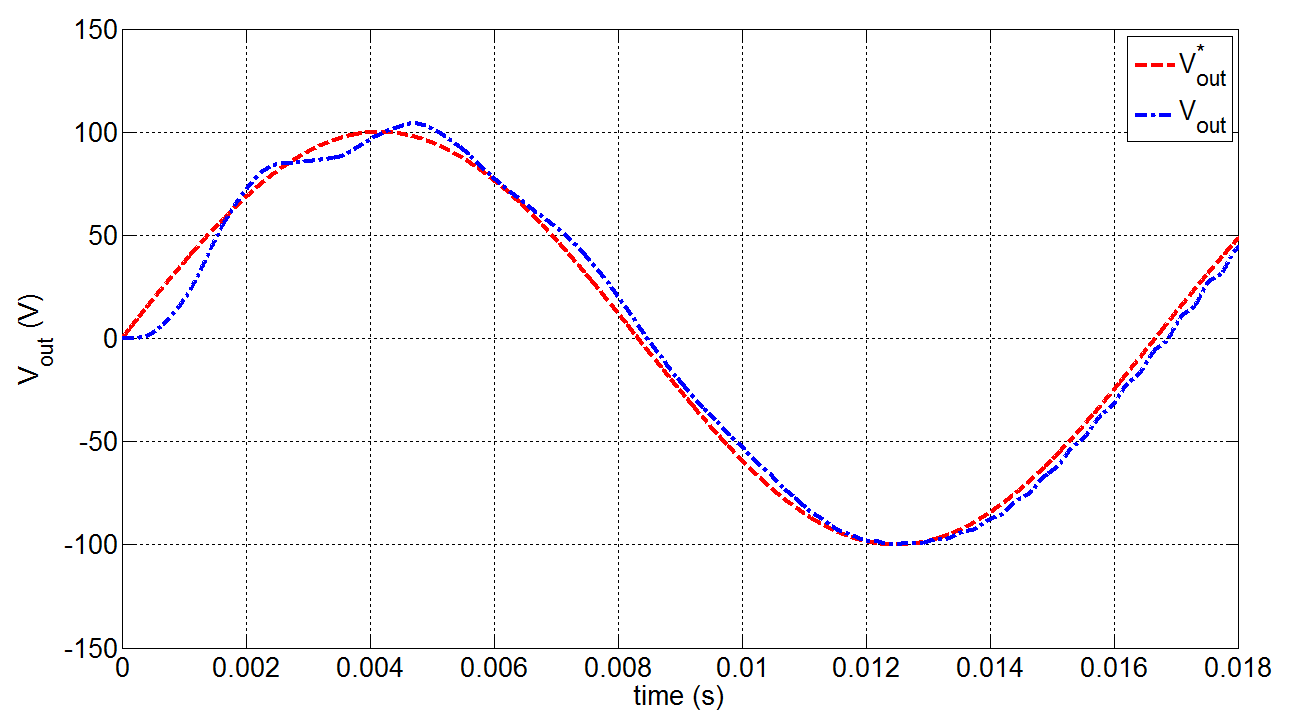}
\caption{Output voltage of the parallelized inverters.}
\label{fig:fig_17}
\end{figure}

Consider two single-phase inverters connected in parallel and working in stand-alone mode (Fig. \ref{fig:fig_16}). According to the power sharing methodology \cite{Lai}, the inverter "1" is controlling the output voltage $v_{out}$ and the inverter "2" is controlling the current $i_{L2}$ Two identical ultra-local models are associated to these inverters with the same parameters $\alpha = 30$, $K_p = 20$ and $K_i = 100$. Figure \ref{fig:fig_17} shows the output controlled voltage $v_{out}$ of the associated inverters.

\section{Concluding remarks}


We presented the model-free control methodology in an electrical network environment. Simulations show encouraging results and show that the model-free control has the following features :

\begin{itemize}
\item robust to strong load / topological load changes (e.g. strong change of the resistor value or addition of a load that may increase the order of the whole system);
\item robust to external perturbations (e.g. grid sinusoidal perturbation);
\item direct control in $abc$ frame for tri-phase systems and non-linear control (e.g. power control).
\end{itemize}

A combination of the proposed control strategies allows to extend the results to the control of multiple sources considering simultaneously voltage, current and power control. Further work concerns the study of the stability of the model-free control in networked systems, and the optimal input-output pairing for MIMO systems.

\section*{Acknowledgements} The article presents results of the project G.0717.11 of the Research Council Flanders (FWO).

%

\end{document}